\documentclass[sigconf,natbib=false]{acmart}
\usepackage{amsmath}

\makeatletter
\def\@ACM@checkaffil{
    \if@ACM@instpresent\else
    \ClassWarningNoLine{\@classname}{No institution present for an affiliation}%
    \fi
    \if@ACM@citypresent\else
    \ClassWarningNoLine{\@classname}{No city present for an affiliation}%
    \fi
    \if@ACM@countrypresent\else
        \ClassWarningNoLine{\@classname}{No country present for an affiliation}%
    \fi
}
\makeatother

\AtBeginDocument{%
  }

\setcopyright{acmcopyright}
\copyrightyear{2018}
\acmYear{2018}
\acmDOI{XXXXXXX.XXXXXXX}

\RequirePackage[
  datamodel=acmdatamodel,
  style=acmnumeric,
  ]{biblatex}

\usepackage{tikz,subcaption}
\begin{document}

\title{Triangle Fees}


\author{Rithvik Rao}
\email{rrao@jumpcrypto.com}
\affiliation{%
  \institution{Jump Crypto}
}

\author{Nihar Shah}
\email{nshah@jumpcrypto.com}
\affiliation{%
  \institution{Jump Crypto}
}

\renewcommand{\shortauthors}{Rao and Shah}

\begin{abstract}
Triangle fees are a novel fee structure for AMMs, in which marginal fees are decreasing in a trade's size. That decline is proportional to the movement in the AMM's implied price, i.e. for every basis point the trade moves the ratio of assets, the marginal fee declines by a basis point. These fees create incentives that protect against price staleness, while still allowing the AMM to earn meaningful fee revenue. Triangle fees can strictly improve the Pareto frontier of price accuracy versus losses generated by the status quo of constant fee mechanisms.
\end{abstract}


\keywords{Automated market maker, Loss versus rebalancing, Decentralized exchange}

\settopmatter{printfolios=true}
\maketitle

\section{Introduction}

When traders make swaps on AMMs, they traditionally pay fees that are constant with respect to the volume traded. A trade that is double the size of another pays twice the fees. 

We introduce the concept of ``triangle fees," or fees that are \textit{declining} on the margin with respect to the volume swapped. A trade that is double size of another pays less than twice the fees. More concretely, rather than charging twenty basis points on the first dollar and second dollar alike of a two-dollar trade, this paper proposes a model in which fees are (say) twenty basis points on the first dollar and nineteen basis points on the second dollar. This example also illustrates the nomenclature: on the graph of volume ($x$) versus marginal fee ($y$), the total fee bill is the area under a downwards-sloped line. This is a trapezoid, or---in the limit---a triangle.

Triangle fees are critical for harnessing the value that arbitrageurs provide to an AMM. In general, arbitrageurs represent both opportunity and risk for the AMM. On one hand, exhausting arbitrage opportunities improves price tracking of the AMM to the market price. On they other hand, they represent a source of loss, as they earn profits at the expense of the AMM's LPs. Constant fees balance this tradeoff poorly: if a trader trades until guaranteed arbitrage profit no longer exists, they will earn high profits but the resulting AMM price will still be incongruent with the true price. Triangle fees can balance this tradeoff well.

To understand the mechanism, consider a stylized example in which the true price between two assets has deviated by 25 basis points from the price implied by the AMM. If the AMM charges constant fees of (say) 20 basis points on swaps, arbitrage profits can only be made until the point that the price deviation is 20 basis points. However, no arbitrage profit can be made beyond this point, as the arbitrageur's marginal profits on swaps are below 20 basis points and so below the fees she pays.

If the AMM charges fees that \textit{decline} on the margin, however, guaranteed arbitrage profit exists past this boundary. In particular, we structure the mechanism such that for every basis point change in the implied price by the arbitrageur's swap, the fees for the next bit of volume decline by a basis point. Thus, once the AMM's implied price moves one basis point (and thus the deviation falls from 25 basis points to 24 basis points), the fees drop on the margin from 20 to 19 basis points. Continuing the example, when the arbitrageur's trade results in the AMM price moving by five basis points and leaves a remaining price deviation of 20 basis points, the fee for the next sliver of volume is 15 basis points. As such, the arbitrageur is incentivized to continue trading since it remains profitable. And when the deviation is 15 basis points, the marginal fees are 10 basis points, and so on---and so the resulting price deviation is zero.

This result is in stark contrast to the constant fee model, which cannot ever hope to correct price deviations to zero---unless the constant fee is set to zero and so the AMM collects no fees. By contrast, this model allows the AMM to collect meaningful fees from arbitrageurs and non-arbitraguers while still keeping deviations from the true price close to zero.

Triangle fees can improve the Pareto frontier that constant fees create, in minimizing both price deviations and losses. High constant fees offer high deviation and low losses. Low constant fees offer low deviation and high losses. Triangle fees can offer strict improvements on both fronts, with lower deviations and lower losses.

Triangle fees can also be beneficial to noise traders, particularly if they place large trades. Triangle fees act as ``volume discounts," as larger trades move the AMM's implied price more; and that can lead to lower fee bills in some cases. Of course, smaller noise traders do not enjoy this benefit as acutely, leading to some tradeoff.

The paper proceeds as follows. Section \ref{sec:motivation} discusses the motivation for triangle fees at a more foundational level. Section \ref{sec:formula} sets up the equations to compute the fees. Section \ref{sec:simulation} tests the idea and showcases the improvement of the Pareto frontier. Section \ref{sec:uninformed} discusses considerations for noise traders and other non-arbitrageur participants. Section \ref{sec:conclusion} concludes.

\section{Motivation}
\label{sec:motivation}
To motivate the design for triangle fees, first consider a model for an AMM that always knew (perhaps through an oracle) the true price of the two assets in its pool. That AMM would compute the total profit on any trade request, and charge exactly that (minus some $\epsilon$) as a lump-sum fee. A trader exhausting an arbitrage opportunity would thus result in the imbalance in the AMM relative to the truce price being corrected, and the trader earns virtually no profits in the process.

Triangle fees are the oracle-free analog to this model. It guesses at the arbitrageur's implied profit function, and sets the fees to maximize revenue capture and price tracking. The AMM does not know the true price, or more precisely the price deviation. But it knows that once an arbitrageur initiates a trade, the price deviation at that point in time must exceed the fee. Moreover, once the implied price in the AMM falls by one basis point, it also knows that the arbitrageur's marginal profit on volume must have fallen by one basis point too.

This motivates the mechanism. An AMM sets some initial fee, and arbitrageurs are not incentivized to trade if the deviation does not exceed this value. Once the deviation crosses that value, arbitrageurs engage because they earn guaranteed gross profits on infinitesimal volume, in proportion to the deviation. As that deviation falls, those marginal gross profits shrink. The triangle fee mechanism thus shrinks fees commensurately, to keep net profits constant. For every basis point the AMM's implied price moves (and thus shrinks the deviation), the fees on marginal volume fall by exactly one basis point. (Note that the exposition discusses the arbitrageur's trade as a series of incremental steps for expositional reasons, but of course the arbitrageur makes the entire swap in a single trade in practice.)

\begin{figure}[!h]
    \centering
    \tikzset{every picture/.style={line width=0.75pt}} 

    \begin{tikzpicture}[x=0.75pt,y=0.75pt,yscale=-1,xscale=1]
    
    \draw  (51.88,200.6) -- (260,200.6)(72.69,31) -- (72.69,219.45) (253,195.6) -- (260,200.6) -- (253,205.6) (67.69,38) -- (72.69,31) -- (77.69,38)  ;
    \draw  [dash pattern={on 4.5pt off 4.5pt}]  (73.13,97.21) -- (257.07,98.06) ;
    \draw    (73.13,47.13) -- (139.09,96.36) ;
    \draw    (139.09,96.36) -- (139.82,200.77) ;
    
    \draw (-1.24,89.14) node [anchor=north west][inner sep=0.75pt]   [align=left] {Marginal Fee};
    \draw (74.38,130) node [anchor=north west][inner sep=0.75pt]   [align=left] {\begin{minipage}[lt]{43.55pt}\setlength\topsep{0pt}
    \begin{center}
    Total\\Fee\\Revenue
    \end{center}
    
    \end{minipage}};
    \draw (72.59,66.08) node [anchor=north west][inner sep=0.75pt]   [align=left] {\begin{minipage}[lt]{26.53pt}\setlength\topsep{0pt}
    \begin{center}
    Net\\Profit
    \end{center}
    
    \end{minipage}};
    \draw (131.4,202.03) node [anchor=north west][inner sep=0.75pt]   [align=left] {Trade Size};
    \draw (14,44.14) node [anchor=north west][inner sep=0.75pt]   [align=left] {Deviation};

    \end{tikzpicture}
    \caption{Arbitrageurs under constant fees. Arbitrageurs are incentivized to trade until the marginal fee boundary is reached, as doing so exceeds the marginal cost.}
    \label{fig:graph1}
\end{figure}
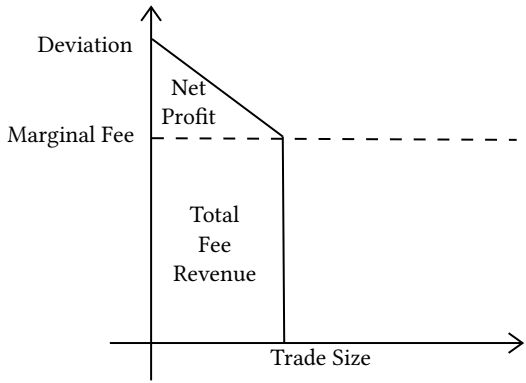

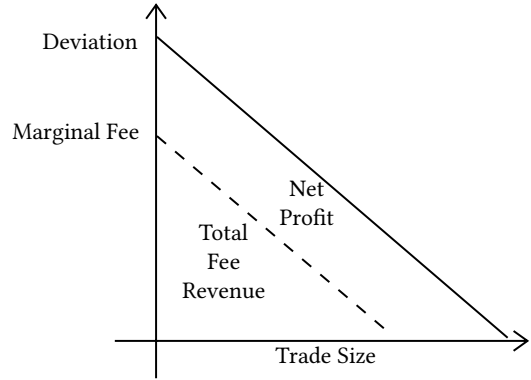
\begin{figure}[!h]
    \centering

    \tikzset{every picture/.style={line width=0.75pt}} 
    
    \begin{tikzpicture}[x=0.75pt,y=0.75pt,yscale=-1,xscale=1]
    
    \draw  (51.88,200.6) -- (260,200.6)(72.69,31) -- (72.69,219.45) (253,195.6) -- (260,200.6) -- (253,205.6) (67.69,38) -- (72.69,31) -- (77.69,38)  ;
    \draw  [dash pattern={on 4.5pt off 4.5pt}]  (73.13,97.21) -- (192,198) ;
    \draw    (73.13,47.13) -- (250,199) ;
    
    \draw (-1.24,89.14) node [anchor=north west][inner sep=0.75pt]   [align=left] {Marginal Fee};
    \draw (76.38,140) node [anchor=north west][inner sep=0.75pt]   [align=left] {\begin{minipage}[lt]{43.55pt}\setlength\topsep{0pt}
    \begin{center}
    Total\\Fee\\Revenue
    \end{center}
    
    \end{minipage}};
    \draw (129.59,118.08) node [anchor=north west][inner sep=0.75pt]   [align=left] {\begin{minipage}[lt]{26.53pt}\setlength\topsep{0pt}
    \begin{center}
    Net\\Profit
    \end{center}
    
    \end{minipage}};
    \draw (131.4,202.03) node [anchor=north west][inner sep=0.75pt]   [align=left] {Trade Size};
    \draw (14,44.14) node [anchor=north west][inner sep=0.75pt]   [align=left] {Deviation};

    \end{tikzpicture}
    \caption{Arbitrageurs under triangle fees. Arbitrageurs' marginal cost per unit of volume decreases with the marginal profit, and so the resulting price deviation will be zero.}
    \label{fig:graph2}
\end{figure}

Graphically, the difference between constant fees and triangle fees can be shown in Figures \ref{fig:graph1} and \ref{fig:graph2}. Figure \ref{fig:graph1} demonstrates that, for a given deviation, an arbitrage opportunity under constant fees exists until the point that the deviation reaches the marginal fee, resulting in net profits after paying fees to the AMM.

By contrast, Figure \ref{fig:graph2} demonstrates that, under triangle fees, the arbitrageur's marginal fees decline with the marginal deviation. Thus, the arbitrageur trades until the point of no price deviation, earning net profits but continuing to pay fee revenue to the AMM.

While triangle fees strongly improve an AMM's price tracking, there is a more complex relationship with respect to losses. All else equal, arbitrageurs make higher profits for a given trade under triangle fees, as they monetize a larger trade. However, arbitrageurs have \textit{fewer} opportunities to make profits under triangle fees, as the deviation crosses the fee boundary less frequently. By contrast, under constant fees, some price deviation must persist because arbitrage opportunities are exhausted prior to reaching the point of no deviation. The price deviation thus constantly hovers at the fee boundary and frequently crosses it, giving arbitrageurs an opportunity to earn small profits by making small trades. As the next sections illustrate, this leads to comparable losses in practice under either regime.

Returning to an earlier point, triangle fees can proxy for an oracle integration. While an oracle integration can make tracking perfectly precise and losses extremely small, an oracle integration also requires substantial engineering work and introduces a dependency on an external service. Many AMMs utilize simple smart contracts to manage the protocol. Various other fee proposals have been suggested with the aim of mitigating LP losses to arbitrageurs, such as fees that adjust based on market volatility \cite{hydraswap2022} and fee mechanisms that aim to directly discriminate toxic flow \cite{crocswap2023}. But these solutions are complex and have not yet been operationalized at scale, and the triangle fee mechanism represents an easier operational change that still achieves similar aims.

Finally, in a direct sense, the motivation for triangle fees is to better manage the AMM's relationship with arbitrageurs. But indirectly, triangle fees can improve the experience for uninformed traders. Superficially, triangle fees are effectively ``volume discounts," which benefits large traders but are less exciting to small traders. But more subtly and importantly, triangle fees---by allowing better price tracking---can improve the total slippage that the uninformed trader faces. Regardless of the fee she pays, that trader executes her trade at a price that is less stale---and that is beneficial for her in general.

\section{Formula}
\label{sec:formula}
AMMs must compute fees for a given trade. This section sets up the framework for a standard constant-product AMM (i.e. $xy = k$) to do so under triangle fees.

Figure \ref{fig:graph2} illustrates that the total fee paid is the area under the curve of marginal fees against trade size. This means that the total fee is simply the integral of that curve. The marginal fee is designed to fall in proportion to the movement in the AMM price; and in turn, the movement in AMM price is a function of the trade size.

Assume an arbitrageur makes a trade to take $\Delta x$ and deposit $\Delta y$ (where $\Delta x$ and $\Delta y$ can be positive or negative). Per the constant product rules, this must obey the constraint that $(x - \Delta x) (y + \Delta y) = xy$. This allows us to link the movement in the AMM price (which itself is just the ratio of the two assets in the pool) to the amount of $\Delta x$ that the trader demands.
\begin{align*}
\Delta p =& \frac{y + \Delta y}{x - \Delta x} - \frac{y}{x} \\
=& \frac{xy}{(x - \Delta x)^2} - \frac{y}{x}
\end{align*}

This allows the following expression to compute the total fee for a given trade, assuming $\Delta x > 0$. We define $f$ to be the initial fee, e.g. twenty basis points in the opening example, and $m$ to be the slope of fee change with respect to price. Throughout this paper (with a few exceptions), we set $m = -1$, i.e. every one basis point change in price leads to a one basis point decline in fee---but defining this as a parameter allows us to easily generalize the framework.
$$
\text{Fee} = \int_0^{\Delta x} \left(f + m\left(\frac{xy}{(x - w)^2} - \frac{y}{x}\right)\right)dw
$$

There are two complications. First, more trivially, this equation needs to be modified appropriately to handle the case where $\Delta x < 0$. Second, more importantly, AMMs will want (and, in fact, need) to set some ``base" fee $b$. This is the lowest possible marginal fee, and the marginal fee cannot decline beyond this. Implicitly in Figure \ref{fig:graph2}, the base fee is set at $b = 0$, but in practice AMMs may want to set it higher. This parameter allows the triangle fee model to nest the constant fee model, by setting $b = f$.

In total, this yields the following expression for the total fee, as a function of the parameters $(f, b, m)$, the initial liquidity in the pool $(x, y)$ and the trader's desired trade size $\Delta x$.

\begin{equation}
\label{eq:equation1}
\text{Fee} = \begin{cases}\!
\begin{aligned}[c]%
f x(1 - k_u) + m\left(\frac{y}{k_u}\right) \\
- my(2 - k_u) + b(\Delta x - x(1 - k_u))
\end{aligned} & \text{if } \Delta x > x (1 - k_u) \\
f\Delta x + m\left(\frac{xy}{x - \Delta x}\right) - m \frac{y}{x} \Delta x - my & \text{if } \Delta x \in [0, x (1 - k_u)] \\
-f\Delta x + m \left(\frac{xy}{x - \Delta x}\right) - m \frac{y}{x} \Delta x - my & \text{if } \Delta x \in [x(1 - k_\ell), 0) \\
\begin{aligned}[b]%
-fx(1 - k_\ell) + m \left(\frac{y}{k_\ell}\right) \\
- my(2 - k_\ell) - b (\Delta x - x (1 - k_\ell))
\end{aligned} & \text{if } \Delta x < x (1 - k_\ell) \\
\end{cases}
\end{equation}

In turn, $k_u$ and $k_\ell$ reflect the point, in relative terms, where the base fee begins to bind on the trade size, and they are computed as follows.
\begin{align}
k_u =& \sqrt{\frac{y/x}{y/x + (b - f)/m}} \nonumber \\
k_\ell =& \sqrt{\frac{y/x}{y/x - (b - f)/m}}
\label{eq:equation2}
\end{align}

These formulas allow us to operationalize a variety of simulated AMMs in the subsequent sections.

\section{Arbitrageur Analysis}
\label{sec:simulation}
Triangle fees can improve the Pareto frontier created by constant fees, and this section illustrates this advancement by testing the framework against arbitrageurs.

We simulate a variety of constant fee and triangle fee specifications. Initial fees start at two basis points, and climb to fifty basis points. Base fees start at two basis points, and climb to the initial fee. For each specification, we simulate fifty different worlds, each where the true price evolves through 20,000 time steps according to a symmetric Gaussian random walk with a standard deviation of three basis points per time step.

Arbitrageurs get to participate at every time step, and make the profit-maximizing trade if one exists given the fee structure (i.e., they trade until guaranteed profit no longer exists). They are required to pay a small gas fee to trade. To keep AMM liquidity comparable between all worlds, fees are not redeposited into the AMM but are paid to a separate account.

For each parameterization, we compute two core metrics averaged across the fifty worlds: total loss faced by the AMM, and average price deviation of the AMM. The former term is more precisely loss-versus-rebalancing (LVR), as defined and interpreted in Milionis et al.\ (2022), minus trading fees captured. \cite{milionis2022automated} In particular, this can be interpreted as the total profit-and-loss (PNL) of the LPs minus the PNL of a simple rebalancing strategy. The latter term is more precisely the square root of the mean squared deviation between the AMM implied price and the true price.

To establish the Pareto frontier present in the status quo, we plot these values for all AMMs that implement constant fees. This gives a clear tradeoff: low constant fees correspond to the top-left of the graph, where losses are large but deviations are small, while high constant fees correspond to the bottom-right of the graph, where losses are small but deviations are large. 

\begin{figure}[H]
    \centering
    \includegraphics[width=0.4\textwidth]{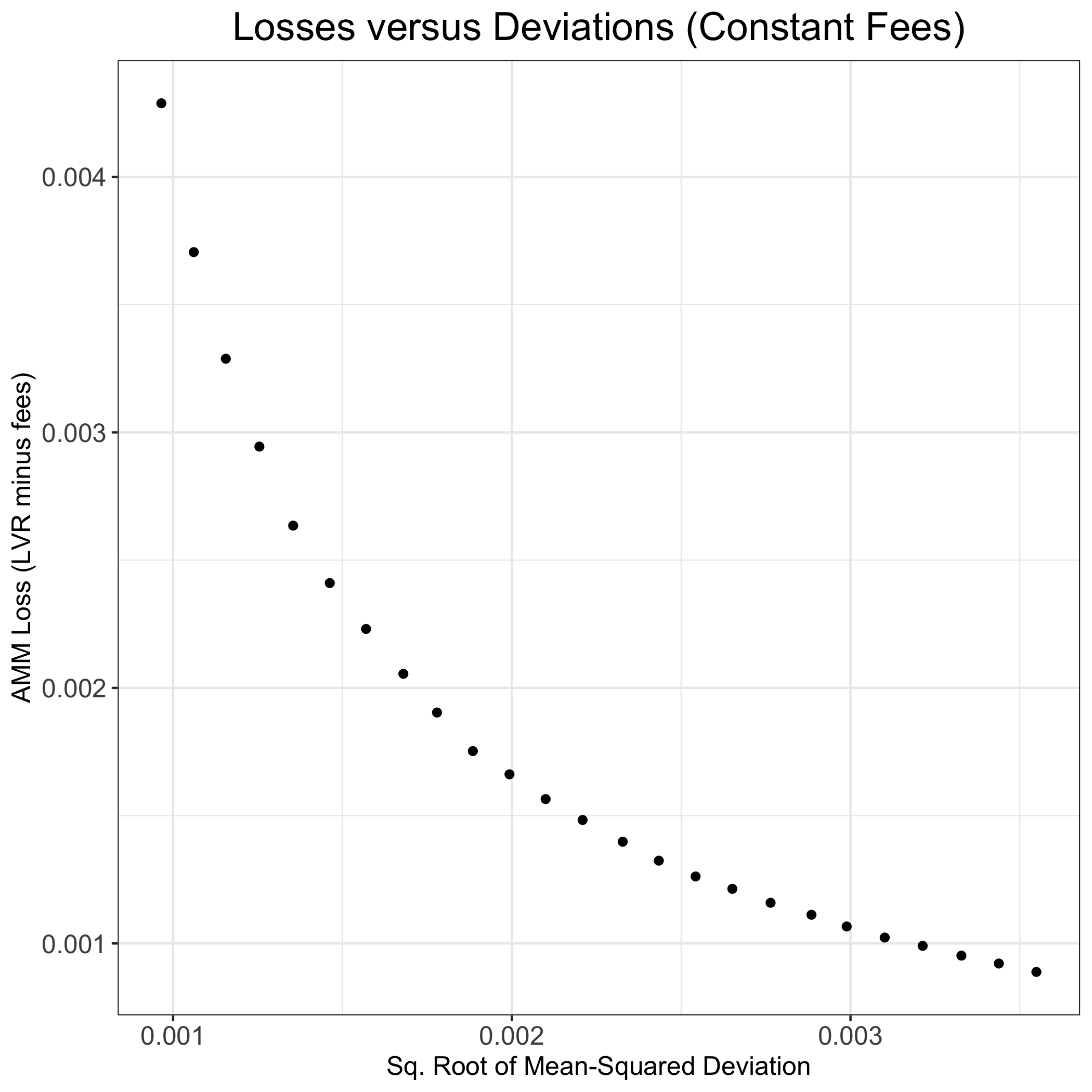}
    \caption{Pareto frontier of constant fees. The top-left and bottom-right extremal points correspond to low fees and high fees respectively. Low constant fees protect against stale AMM prices, while high constant fees protect against high losses.}
    \label{fig:graph3}
\end{figure}

\begin{figure}[H]
    \centering
    \includegraphics[width=0.4\textwidth]{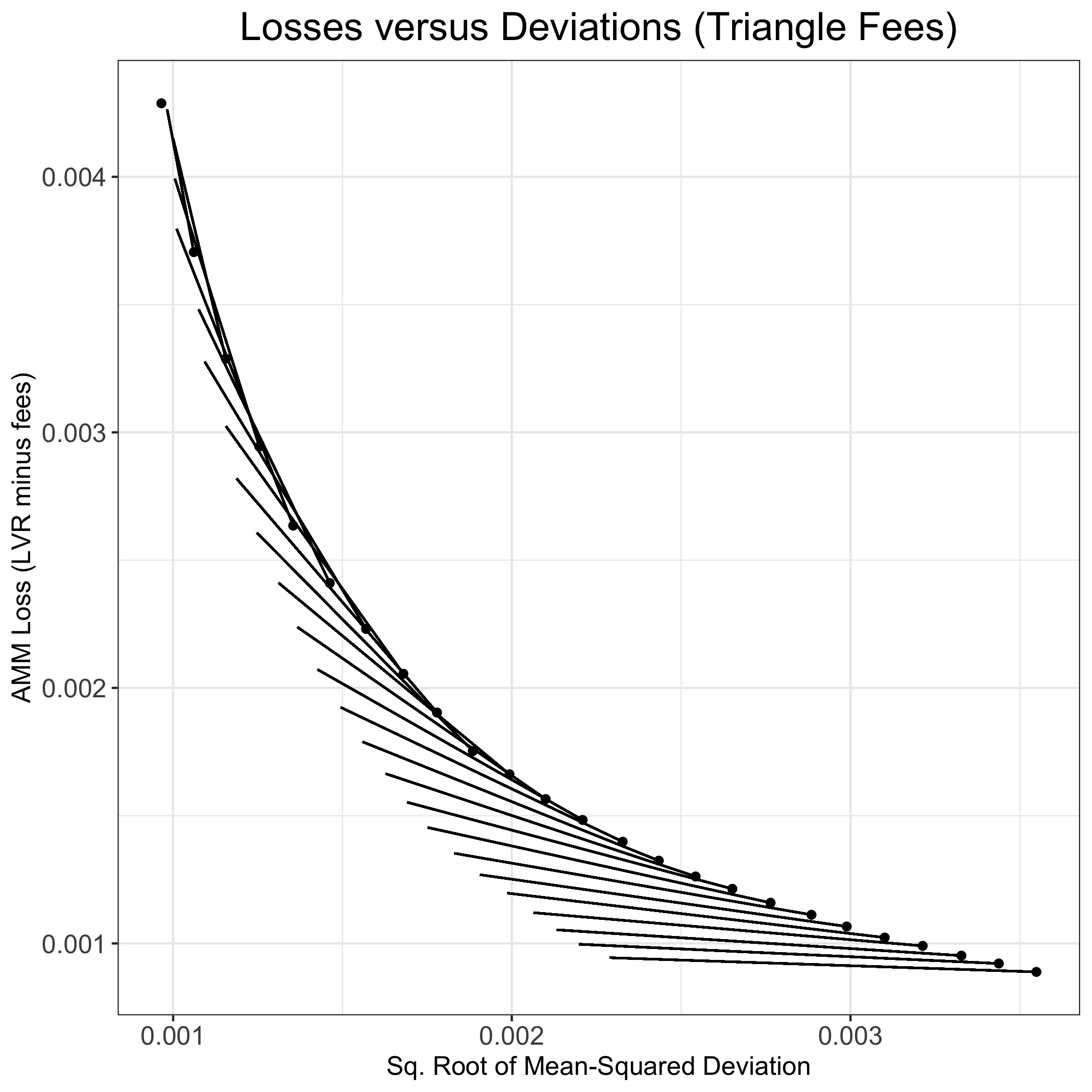}
    \caption{Pareto frontier with triangle fees. Each line reflects the line of best fit for all parameterizations for the same initial fee and differing base fees. For each line, the rightmost point is equivalent to a constant fee parameterization, where the initial and base fees are equivalent; and the leftmost point is equivalent to a parameterization with a base fee of two basis points.}
    \label{fig:graph4}
\end{figure}

We overlay triangle fees in this plot in Figure \ref{fig:graph4}. To make the graph readable, we plot lines rather than points, where each line connects parameterizations that share the same initial fee but differ in their base fee. (For instance, the parameterizations with $(f, b) = (30, 20)$ and $(f, b) = (30, 10)$ basis points would be used to construct the same line.) Each line of best fit terminates in one of the original points on the Pareto frontier, because a triangle fee parameterization where $f = b$ is identical to a constant fee parameterization at $f$.

Figure \ref{fig:graph4} is striking, in its improvements of the the Pareto frontier. For a given constant fee parameterization, there exist \textit{many} triangle fee parameterizations that offer lower losses and lower price deviations. This is the core result of the paper: that triangle fees can break the Pareto frontier.

In alternate simulations that are not shown, triangle fees are tested against environments with higher or lower gas fees, and higher or lower true price volatilities. The results are qualitatitively the same. Quantitatively, triangle fee model tend to outperform constant fee models by more when gas is high or when volatility is low. However, theoretical work to explain this comparative static remains ongoing, and so we do not surface it in the paper.

\begin{figure}[!h]
    \centering
    \includegraphics[width=0.4\textwidth]{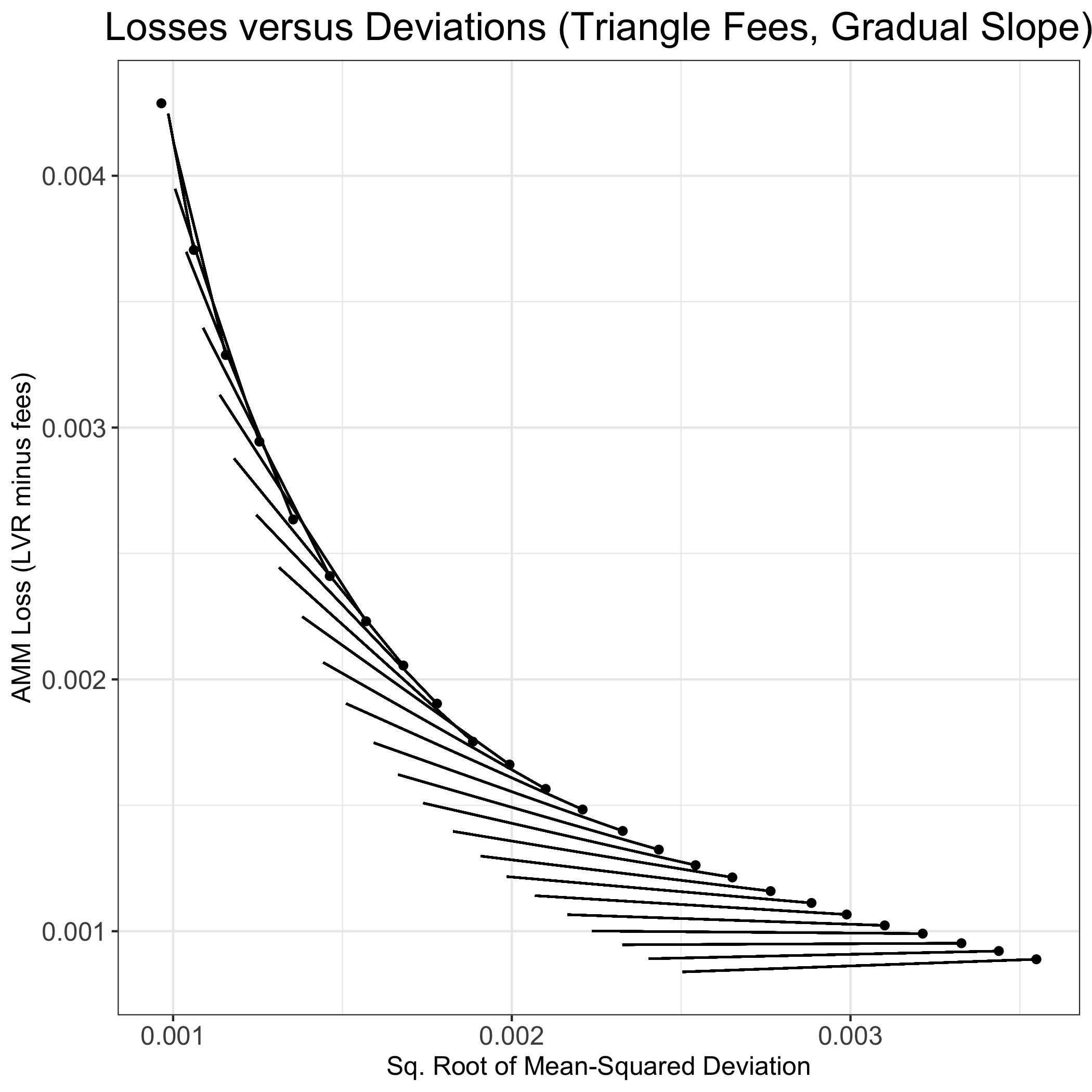}
    \caption{Pareto frontier with shallower triangle fees. This figure admits the same interpretation as Figure \ref{fig:graph4}, except with a slope of $m = -0.8$ rather than a slope of $m = -1$.}
    \label{fig:graph5}
\end{figure}

Finally, the focus of the paper is on the mechanism with a slope of $m = -1$. However, we can briefly consider alternate slopes for Equations \eqref{eq:equation1} and \eqref{eq:equation2}. In particular, consider a shallower slope of $m = -0.8$ (i.e. the fee declines by 0.8 basis points for every basis point change in the price). Figure \ref{fig:graph5} presents the same Pareto frontier with this slope.

Relative to Figure \ref{fig:graph4}, Figure \ref{fig:graph5} is unique in one respect. In general, the Pareto improvements in Figure \ref{fig:graph4} can only be accessed by changing the fee structure meaningfully (e.g. a constant fee of thirty basis points must migrate to an initial and base fee of forty and twenty basis points). But in Figure \ref{fig:graph5}, the changes can be more subtle and still access Pareto improvements in a few special cases. For instance, a constant fee of fifty basis points is outperformed by an initial and base fee of fifty and thirty basis points.

We also consider a framework with a steeper slope, e.g. $m = -1.2$. This offers a different array of options to an AMM designer, but it is qualitatively identical to Figure \ref{fig:graph4}.

\section{Noise Traders}
\label{sec:uninformed}

Triangle fees can Pareto dominate constant fees along the dimensions of losses and mean-squared deviation, but there are more subtle equilibrium effects on noise traders and other non-arbitrageur participants. These are important to understand, to conclusively describe the result of implementing them.

For instance, the analysis of Section \ref{sec:simulation} supposes that an AMM moving from a constant fee to a triangle fee does not alter the composition of trades it faces. This is highly realistic for arbitrageurs, who participate to exhaust any opportunities available, resulting in smaller price deviations. But this is less realistic when discussing noise traders, whether they place trades directly or place trades indirectly through routers with fixed algorithms. For this class of traders, the AMM must provide a welcoming environment for them to execute swaps.

As one example, trades with large notional sizes (made by so-called ``whales'') may be more attracted to the AMM, due to the effective volume discount provided by triangle fees. This is equally beneficial for the AMM, as such trades still generally generate high fees and do not typically contribute to LVR. As a second example, trades that go through routing algorithms may also favor triangle fees, due to the lower mean-squared price deviation. These routers typically maximize overall return, and constant-fee AMMs would look less appealing to them by presenting stale prices. As a final example, traders with $\epsilon$-sized trades could face higher fees in an AMM with triangle fees, where initial fees can be high. As a result, they might be deterred by the high salience of fees and choose to execute swaps at other venues.

A complete understanding of the outcome of implementing triangle fees would require characterizing the elasticity of the distribution of noise trades with respect to the fee regime. This elasticity is difficult, if not impossible, to study empirically with rigor. Each of the effects suggested above are driven by different factors, with differences in simplicity, salience, and loss. Finally, a \textit{full} analysis of this question would also add liquidity providers into the mix. LPs, who would face lower losses under triangle fees, may choose to provide liquidity to the AMM more enthusiastically, which would in turn improve the AMM's attractiveness to all traders at large.

However, we perform some analysis on the question. In particular, we consider how the ``median" noise trader, concerned about total slippage, might fare under triangle fees.

To calibrate this analysis with actual data, we study trades made on Uniswap. In particular, we consider all trades made in the Uniswap v3 USDC/ETH pool with a five basis point constant fee, in the interval between January 1, 2023 and March 31, 2023, using data sourced from Bitquery's GraphQL Ethereum API. This pool is the most liquid AMM-based decentralized exchange, and daily trading volume regularly exceeds hundreds of millions of dollars. \cite{uniswapusdceth}

When analyzing flow from noise traders, we restrict ourselves to swaps initiated by smart contracts known to typically mediate non-arbitrage trades. This analysis is permitted by the fact that successful arbitrageurs usually interact with AMM contracts directly and avoid use of routers or other front-end interfaces. We can do this by parsing the \texttt{to\_addr} specified by each swap event. The addresses we parse include those of the various Uniswap routers, 0x, 1inch, CoW Protocol, Coinbase, MetaMask, and Paraswap. We find just over 165,000 swaps in this timeframe meet our criteria.

\begin{figure}[!h]
\centering
\includegraphics[width=0.9\linewidth]{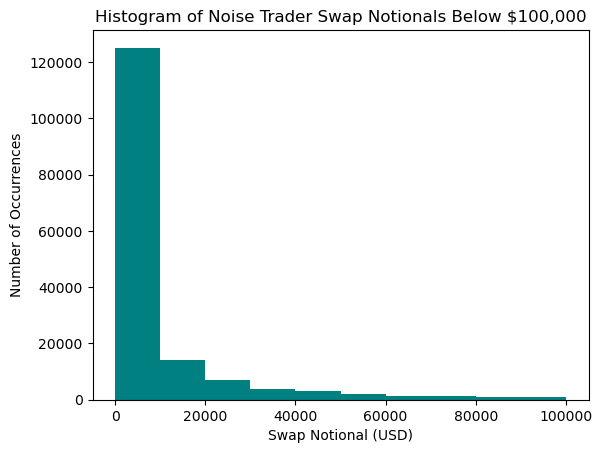} 

\includegraphics[width=0.9\linewidth]{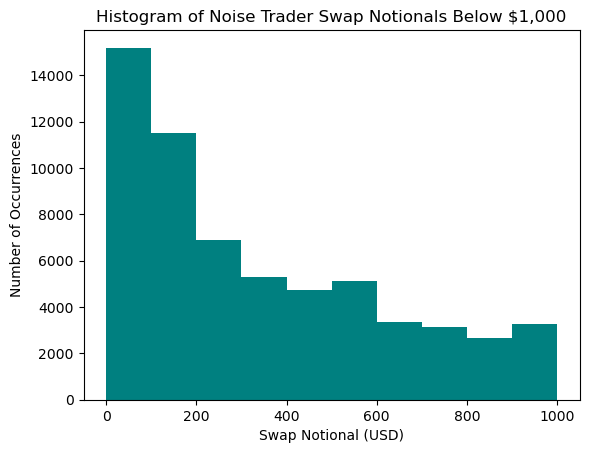}
\caption{The distribution of notional sizes of swaps originated by noise traders in the Uniswap v3 USDC/ETH five-basis point pool in Q1 2023. The top histogram shows the distribution of swaps below \$100,000 notional value, while the bottom one shows the distribution below \$1,000.}
\label{fig:histswapntnl1}
\end{figure}

Figure \ref{fig:histswapntnl1} presents the distribution of notional values of swaps made in the pool. It does so by marking each trade to the implied DEX price prior to the occurrence of the trade. The results generate a well-known power law dynamic present in many financial datasets, that is largely invariant to scale. (Stunningly, this result emerges even after \textit{truncating} the most extreme swaps to make the figure readable.) The modal swap is small, but the maximum swaps are large.

To understand these results in terms of triangle fees, we must consider what impact the trades have on the implied AMM price. The price impact (and, consequently, how the fee is computed) is a function of the distribution of notional swap size relative to the pool's available liquidity. In particular, we define \textit{price impact} of a trade as the absolute difference between the implied DEX price before and after the trade, divided by the implied price before the trade. We report price impact in basis points (bps).\footnote{We measure the implied DEX price after a swap by observing the price reported by the next swap in the pool, which is exposed by the \texttt{sqrtPriceX96} field in Uniswap \texttt{Swap} events. This is not an entirely correct measure, since additions and removals of liquidity in the concentrated liquidity model of Uniswap v3 can shift the price between swaps. Assuming that the expected price impact due to changes in liquidity is zero (such as if the distribution of price changes due to LP actions is symmetric about zero), the analysis is robust to this technicality.}

Table \ref{tab:swappriceimpact} presents the quantiles of the distribution of price impact of the swaps under study, and---much like the distribution of notional swap sizes---this distribution exhibits a fat tail. The median swap has a price impact of just over 3.7 basis points. Depending on the triangle fee specification (such as if the initial fee is ten basis points, rather than a constant six basis points), this median swap may face a higher effective fee than in the constant fee model. However, a long tail of sizable swaps with substantial price impact may be attracted to trade on the AMM.

\begin{table}
\begin{tabular}{l|r}
\textbf{Quantile} & \textbf{Price Impact (bps)} \\
\hline
0.05             & 0.0069                 \\
0.25              & 0.1021        \\
0.50             & 3.7774                \\
0.75           & 9.9981               \\
0.95              & 10.7545       \\
0.99 & 17.3149 \\
0.999 & 69.2415 \\
0.9999 & 212.1279
\end{tabular}
\caption{Quantiles of the distribution of price impact of swaps likely originated by noise traders (Uniswap v3 USDC/ETH 5 basis point pool, 01/01/2023--03/31/2023).}
\label{tab:swappriceimpact}
\end{table}

We apply the results in Table \ref{tab:swappriceimpact} to our setting. For each of our simulated AMMs, at each time step, and in each direction (swap $x$ for $y$, and swap $y$ for $x$), we back out the size needed to adjust the AMM's implied price by the various quantiles specified in Table \ref{tab:swappriceimpact}. We then measure the total slippage, inclusive of fees, that this corresponding trade faces. Finally, we compute and report the mean-squared value of that across all time steps and all AMMs.\footnote{For any parameterization of an AMM at any point in time, one swap will enjoy low slippage because they will be implictly correcting an imbalance, while one swap will suffer high slippage because they will be furthering an imbalance. Thus, it is important to measure the impact on swaps in both directions; and to measure the mean squared deviation rather than the mean deviation.}

\begin{figure}[!h]
    \centering
    \includegraphics[width=0.4\textwidth]{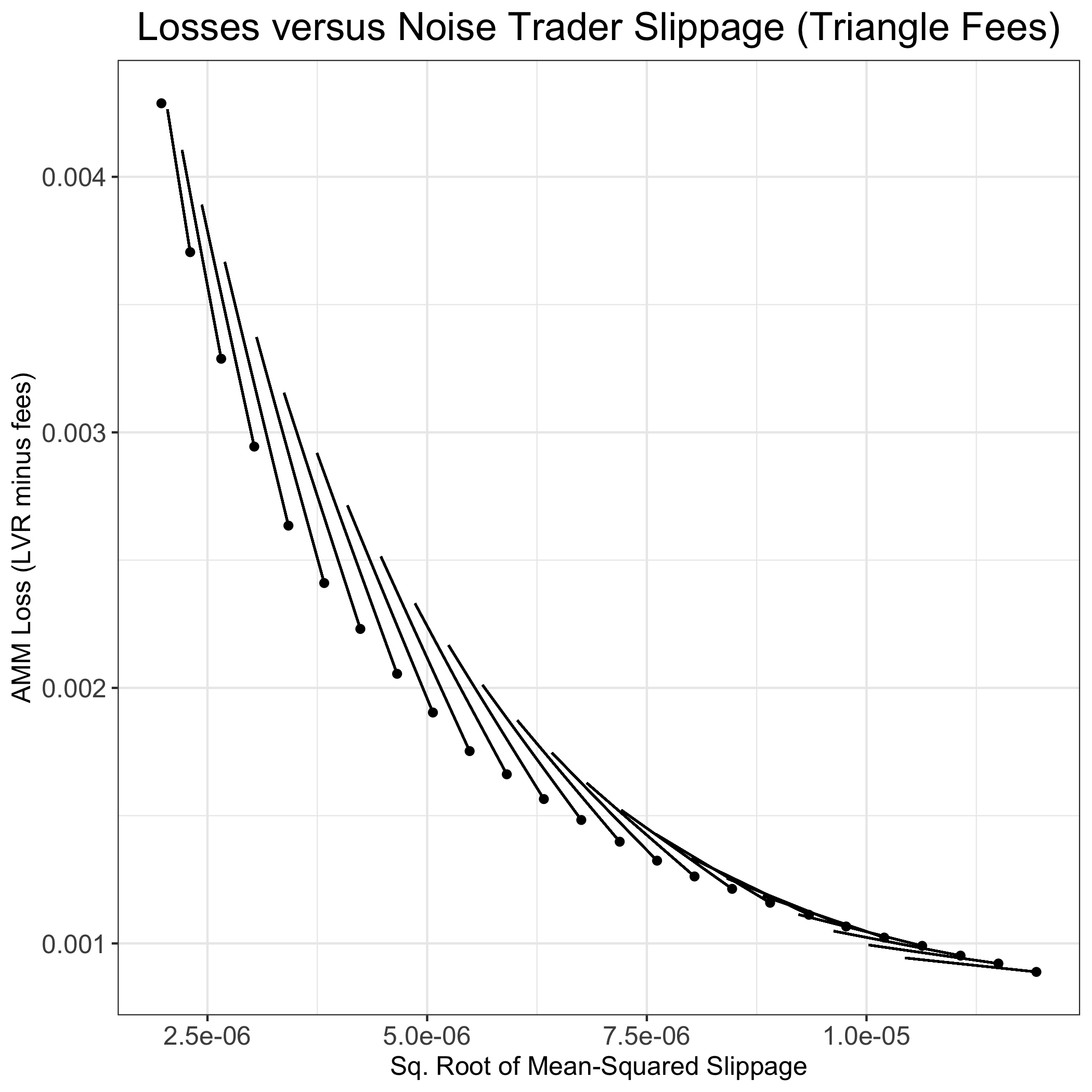}
    \caption{Pareto frontier for noise traders. The $y$-axis continues to represent AMM loss. The $x$-axis now represents the square root of mean-squared slippage rather than that of price deviation, and represents the total slippage a median noise trader would face.}
    \label{fig:graph6}
\end{figure}

Figure \ref{fig:graph6} demonstrates the results, focusing on trades corresponding to the median price movement of 3.8 basis points. The results are not ideal, but they show promise. For a few cases (namely when constant fees are high), triangle fees improve upon the Pareto frontier of slippage and loss. However, for most cases, triangle fees do not improve upon that Pareto frontier. While the enhanced price accuracy that triangle fees provides to the AMM is beneficial, the fees paid by those traders tend to be higher under triangle fees. On net, this tends to increase the total slippage the median noise trader in a pool like Uniswap v3 USDC/ETH faces---and so the lines sometimes extend outwards rather than inwards. In many respects, the impressive depth of this pool makes most swaps seem minuscule in relative terms, and that limits the full potential of triangle fees.

\begin{figure}[!h]
    \centering
    \includegraphics[width=0.4\textwidth]{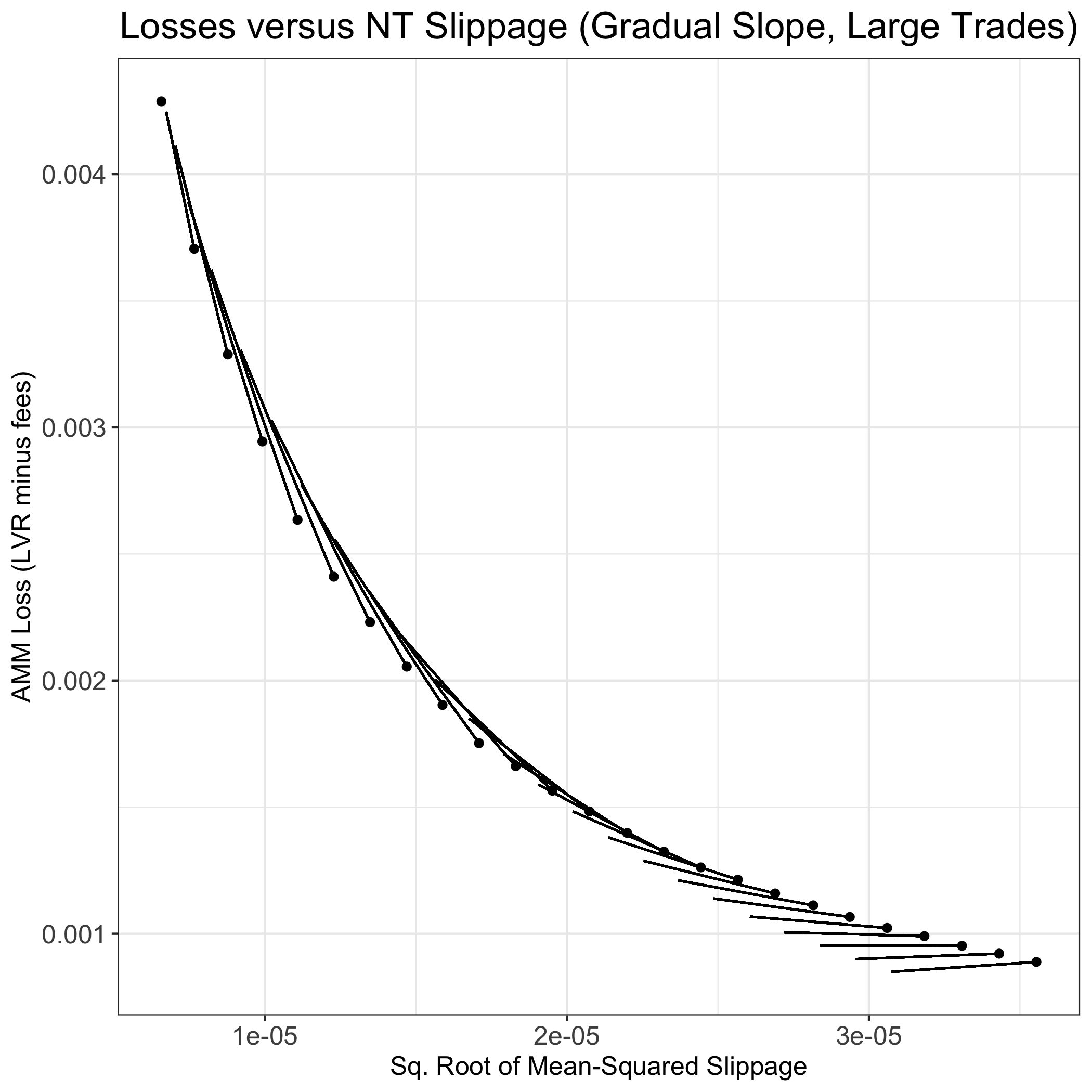}
    \caption{Pareto frontier for larger noise traders and under shallower slopes. This figure admits the same interpretation as Figure \ref{fig:graph6}, except the noise traders place larger trades and the slope of the triangle is $m = -0.8$ rather than a slope of $m = -1$.}
    \label{fig:graph7}
\end{figure}

However, there are cases where this model can improve the frontier for noise traders too. For instance, large trades benefit from this framework, as they capitalize effectively on the ``volume discounts" that the fee structure provides. In addition, the fee structure with a shallower slope for the triangle also sees larger improvements, likely for the same empirical reasons that Figure \ref{fig:graph5} improves on Figure \ref{fig:graph4}. These are both jointly demonstrated in Figure \ref{fig:graph7}, where the slope is $m = -0.8$ and each noise trader alters the AMM implied price by 10.75 basis points (which corresponds the 95th-percentile in Table \ref{tab:swappriceimpact}). Now, the model with triangle fees improves upon the Pareto frontier much more meaningfully.

Thus, triangle fees may not be appropriate for all traders in all cases. But, they show promise in certain circumstances---particularly when trades are large, when pool liquidity is low, or when slopes deviate from the stylized framework. Indeed, future work can explore combinations of triangle and constant fees in tandem, to further improve upon the Pareto frontier.

\section{Conclusion}
\label{sec:conclusion}
Triangle fees operationalize a key economic insight: match the arbitrageur's marginal costs to their marginal revenues. That change allows AMMs to use arbitrageurs in novel ways, relative to the status quo. Arbitrageurs are given incentives that enable the AMM to keep prices less stale, while not earning outsized returns at the AMM's expense. Future work will identify the settings in which they are most useful.

For all their naysayers, simple AMMs have remained enduring features of the DeFi landscape. This simplicity is prudent, especially from a security standpoint. But this simplicity need not be incompatible with innovation. Triangle fees are one such example: they are easy to develop, but they can make AMMs much more robust and useful.


\newpage
\printbibliography



\end{document}